\documentclass[a4paper, oneside, 12pt]{article}
\usepackage{amsmath,amsthm,amssymb}
\usepackage{ bbold }
\usepackage[utf8]{inputenc}
\usepackage[T1]{fontenc}
\usepackage{graphicx}
\usepackage{longtable}
\usepackage[hidelinks]{hyperref}
\usepackage{cleveref}
\usepackage{tikz} 
\usepackage{pgfplots}
\usetikzlibrary{arrows}
\pgfplotsset{compat=1.15}
\usepackage{mathrsfs}
\usepackage{physics}
\usepackage[most]{tcolorbox}
\usepackage{caption}
\usepackage{caption,subcaption}
\usepackage{comment}
\usepackage{float}
\usepackage[most]{tcolorbox}
\usepackage{authblk}
\hypersetup{colorlinks,linkcolor={blue},citecolor={blue},urlcolor={red}}  

\usepackage[left=2.5cm, right=2.5cm, top=2.5cm, bottom=2.5cm, bindingoffset=1.5cm, head=15pt]{geometry} 
\usepackage{setspace}
\usepackage{fancyhdr}
\usepackage{qcircuit}
\usepackage{algorithm,algpseudocode}
\usepackage{enumitem}
\newtheoremstyle{break}
  {\topsep}{\topsep}%
  {\itshape}{}%
  {\bfseries}{}%
  {\newline}{}%
\theoremstyle{break}
 \theoremstyle{theorem}
\newtheorem{thm}{Theorem}[section]

\theoremstyle{definition}
\newtheorem{dfn}[thm]{Definition}

\newtheorem{asm}[thm]{Assumption}

\usetikzlibrary{arrows.meta,calc,decorations.pathreplacing,patterns}

\definecolor{blueA}{RGB}{35,95,220}
\definecolor{redB}{RGB}{220,35,30}
\definecolor{purpleT}{RGB}{115,45,145}
\definecolor{gapcol}{RGB}{240,180,50}
\definecolor{boxline}{RGB}{105,100,135}

\usepackage[
    backend=biber,
        style=ieee,
  ]{biblatex}
\newcommand{\bs}[1]{\boldsymbol{#1}}

\theoremstyle{remark}

\definecolor{ududff}{rgb}{0.30196078431372547,0.30196078431372547,1}

\title{The Overlap Gap Property: Separating Quantum and Quantum-Inspired Approximate Optimization Algorithms}
\author[1,2]{Mark Goh}
\affil[1]{Institute for Frontier Materials on Earth and in Space, German Aerospace Center}
    \affil[2]{Institute for Theoretical Physics, University of Cologne}
\author[3]{Thorge M\"uller}
\affil[3]{Institute of Software Technology, German Aerospace Center}
\date{\today}

\begin{document}

\maketitle

\begin{abstract}
In this paper, we prove that the Mean-Field Approximate Optimization Algorithm (MF-AOA), a quantum-inspired classical algorithm of the Quantum Approximate Optimization Algorithm (QAOA), is unable to give arbitrary near optimal solution for problems exhibiting the Overlap Gap Property (OGP). We show this by relating the MF-AOA to finite-memory Approximate Message Passing (AMP) algorithms, which are known to be limited in performance by the OGP. Thus, our paper argues that problems exhibiting the OGP are one such instance where the QAOA can outperform the MF-AOA and AMP algorithms in general. We supplement this by numerically simulating the performance of the QAOA on instances of Max-$4$-XORSAT that exhibit the OGP and find that the QAOA is able to surpass the barrier, while our finite-size scaling analysis favors a super-polynomial over a polynomial depth dependence. In addition, we observe a non-differentiable point in the performance of the QAOA near by the OGP barrier, where the optimal QAOA parameters change from an adiabatic to a non-adiabatic schedule.
\end{abstract}

\section{Introduction}
\begin{figure}[t]
\centering
\resizebox{0.98\linewidth}{!}{
\input{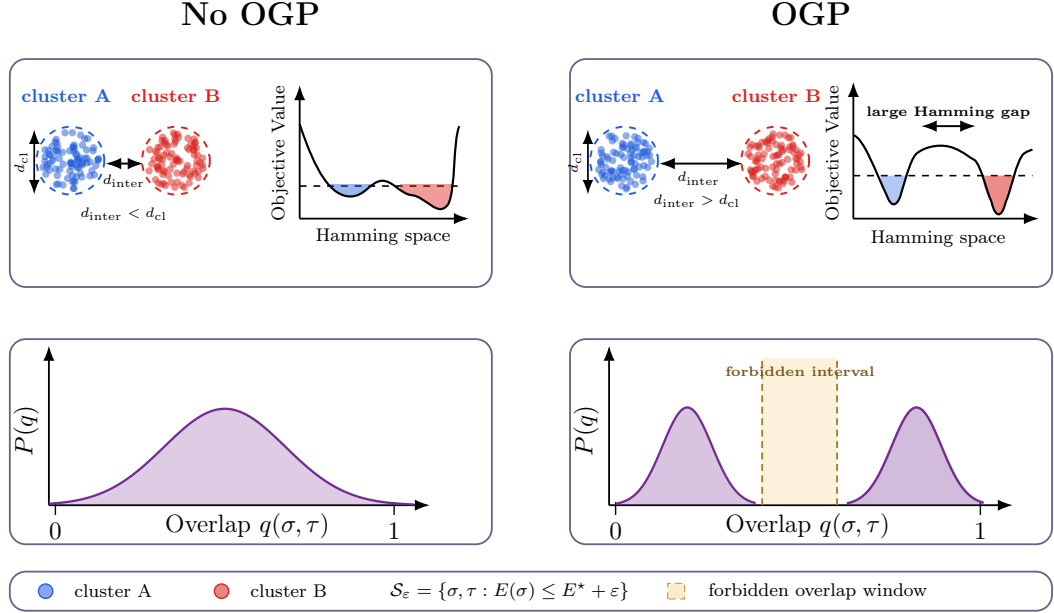}
}
\caption{
Schematic illustration of the overlap gap property (OGP). 
\textbf{Top row:} Geometry of the near-optimal solution space $\mathcal{S}_{\varepsilon}=\{\sigma:E(\sigma)\le E^{\star}+\varepsilon\}$, where $\sigma$ is a spin configuration, $E(\sigma)$ is its objective value or energy, $E^{\star}$ is the optimal ground-state energy, and $\varepsilon$ sets the allowed energy tolerance above the optimum. Without OGP (left), the solution space forms a connected low-energy region in configuration (Hamming) space, where the inter-cluster distance is smaller than the cluster diameter ($d_{\rm inter}<d_{\rm cl}$). Consequently, local modifications of a configuration can continuously connect nearby near-optimal solutions. In the OGP regime (right), the near-optimal solution space fragments into disconnected clusters separated by large Hamming distances, with inter-cluster distance exceeding the cluster diameter ($d_{\rm inter}>d_{\rm cl}$), preventing local transitions between clusters. \\
\textbf{Bottom row:} Illustrative probability distribution $P(q)$ of pairwise overlaps $q(\sigma,\tau)=N^{-1}\sum_i\sigma_i\tau_i$ between uniformly sampled near-optimal configurations $\sigma,\tau\in\mathcal{S}_{\varepsilon}$. Without OGP (left), all intermediate overlap values occur, giving rise to a connected overlap distribution. In contrast, the OGP regime (right) exhibits a forbidden interval of overlaps in which $P(q)=0$, reflecting the absence of near-optimal configurations connecting the separated clusters.
}

\label{fig:ogp_scheme}
\end{figure}

The Overlap Gap Property (OGP) is a topological property of the near-optimal solutions in an optimization problem that limits the performance of various algorithms from finding solutions arbitrarily close to the global optimum \cite{CGPR19,GJ21}. The geometric structure associated with the overlap gap property is illustrated schematically in Fig.~\ref{fig:ogp_scheme}.

Currently, it is known that for sparse graphs, the OGP limits the performance of the Quantum Approximate Optimization Algorithm (QAOA) when the number of layers $p$ used in the QAOA is of logarithmic-depth compared to the problem size $N$ (i.e.\ $p\sim \mathcal{O}(\log N)$) \cite{FGG20, chou_et_al:LIPIcs.ICALP.2022.41}. However, it is known that for any finite-size problem, the QAOA is able to find exact solutions via reduction to the Quantum Adiabatic Algorithm \cite{farhi2014quantumapproximateoptimizationalgorithm}. Thus, it is an open question as to when the expectation value of the QAOA surpasses the OGP barrier. To the best of the authors' knowledge, there has been no explicit study to identify the depth required to surpass the OGP threshold. The closest example is a work by Boulebnane and Montanaro \cite{QAOA_SAT} where they found that constant-depth QAOA had an exponential scaling to find the optimal solution for $k$-SAT near the satisfiability threshold, another problem known to exhibit the OGP \cite{OGP_origin}. \\
In another line of work, a quantum-inspired classical algorithm of the QAOA was created known as the Mean-Field Approximation Optimization Algorithm (MF-AOA) \cite{MFAOA}. The authors also provided a heuristic argument for optimization problems for which the QAOA might still have a quantum advantage. A similar follow-up work showed that the QAOA might have an advantage over the MF-AOA but requires large circuit depth $p$ for the Max-$q$-XORSAT problem \cite{XOR_QAOA}. While no analysis was done with respect to the OGP, it is worth noting that it appears that the MF-AOA similarly was unable to surpass the OGP barrier when comparing their numerical analysis to ours.\\
In this paper, we prove that the MF-AOA is unable to surpass the OGP barrier in general by relating it to Approximate Message Passing (AMP) algorithms (Sec.~\ref{sec:theory_arg}) which have been proven to be limited in performance when the OGP is present. In addition, we perform a numerical analysis (Sec.~\ref{sec:num_arg}) to investigate the circuit depth required for the QAOA to surpass the OGP barrier by classically simulating the QAOA performance on the Max-$4$-XORSAT problem. Our finite-size numerical results favor a super-polynomial over a polynomial scaling of the required QAOA depth and thereby achieve quantum advantage. In addition, we observe that the QAOA approximation ratio develops a non-differentiable point near the OGP barrier, accompanied by a qualitative reorganization of the optimal variational parameters from an adiabatic-like to a non-adiabatic schedule.

\subsection{Notation}

Denote the binary list $\{-1,1\}^N = \mathcal{B}^N$ and the Hilbert cube $[-1,1]^N = \mathcal{H}^N$.\\

Given an $N-$tensor array $J=(J_{i_1,\dots,i_q},1 \le i_1,\dots,i_q \le N)\in (\mathbb{R}^N)^{\otimes q}$ of order $q$ and an $N$-vector $u\in \mathbb{R}^N$, define the inner tensor product as 
\begin{align}
    \label{eq:tensor}
    J(u)= \sum_{1 \le i_1,\dots,i_q \le N} J_{i_1,\dots,i_q} u_{i_1}\dots u_{i_q}.
\end{align}
 Let $\|J\|_2 $ denote the Frobenius norm
\begin{align}
    \| J \|_2 = \sqrt{\sum_{1\le i_1,\dots, i_q \le N} J_{i_1,\dots, i_q}^2}. 
\end{align}
Now for any $u\in \mathbb{R}^N$, denote
\begin{align}
    y = J(\cdot,u)\in \mathbb{R}^N
\end{align}
with 
\begin{align}
    \label{eq:Tensor_Component}
    y_i = \sum_{1 \le i_1,\dots, i_{q-1}\le N } J_{i,i_1,\dots, i_{q-1} } u_{i_1}\dots u_{i_{q-1}}.
\end{align}

\subsection{Related Works}
While the focus of this paper is on the depth and time complexity required of the QAOA to surpass the OGP barrier, we have decided to write a separate paper regarding parameter optimization in the presence and absence of the OGP. A closely related numerical work \cite{WL24} studies how the approximation ratio, the quality of the solution produced by the QAOA, varies between MaxCut and Maximum Independent Set (MIS) while varying the degree of the underlying graph. In that work, they found that for MaxCut, a problem unlikely to exhibit the OGP, the approximation ratio improves for MaxCut as the degree increases whereas the opposite effect was observed in MIS which is known to exhibit the OGP for sufficiently large degree.\\
With regard to quantum advantage and the QAOA, Ref.~\cite{QAOA_JPM_SK} found numerical evidence that the QAOA offers a speed-up in $\epsilon$ for solving the Sherrington--Kirkpatrick problem. Similarly, Ref.~\cite{QAOA_SAT} found that the QAOA has at best a quadratic speed-up in solving $k$-SAT problems at the satisfiability limit while Ref.~\cite{QAOA_LABS} found that the QAOA, combined with quantum minimum finding, gives the best empirical scaling of any algorithm for the Low Autocorrelation Binary Sequence Problem.  For shallow-depth quantum advantage, Ref.~\cite{MZ25} found that the QAOA can find the optimal solution for (near-)symmetric optimization problems requiring only a single layer $p=1$ with at most $\mathcal{O}(\sqrt{n})$ measurements beating general-purpose SAT solvers in some instances when the symmetry is broken.

\section{Background}

\subsection{QAOA}

The QAOA is a quantum algorithm designed to find approximate solutions to combinatorial optimization problems \cite{farhi2014quantumapproximateoptimizationalgorithm}. The goal is to find a bit string $\bs{z}\in \mathcal{B}^N$ that maximizes the cost function $C(\bs{z})$. Given a classical cost function $C$, we can define a corresponding quantum operator $\hat{C}$ that is diagonal in the computational basis, $\hat{C}\ket{\bs{z}}=C(\bs{z})\ket{\bs{z}}$. In addition, define the operator $\hat{B}=\sum_{j=1}^N \hat{X}_j$, where $\hat{X}_j$ is the Pauli $\hat{X}$ operator acting on qubit $j$. Given a set of parameters $\bs{\gamma}=(\gamma_1,\dots,\gamma_p)\in \mathbb{R}^p$ and $\bs{\beta}=(\beta_1,\dots,\beta_p)\in \mathbb{R}^p$, the QAOA prepares the initial state as $\ket{s}=\ket{+}^{N}=2^{-N/2}\sum_{\bs{z}}\ket{\bs{z}}$ and applies $p$ layers of alternating unitary operators $e^{-i\gamma_k \hat{C}}$ and $e^{-i\beta_k \hat{B}}$ to prepare the state
\begin{align}
    \ket{\bs{\gamma,\beta}}=e^{-i\beta_p \hat{B}}e^{-i\gamma_p \hat{C}} \dots
    e^{-i\beta_1 \hat{B}}e^{-i\gamma_1 \hat{C}} \ket{s}.
\end{align}
For a given cost function $C$, the corresponding QAOA objective function is the expectation value $\expval{\hat{C}}{\bs{\gamma,\beta}}$. Preparing the state $\ket{\bs{\gamma,\beta}}$ and measuring in the computational basis yields a distribution over bit strings. Heuristic strategies to optimize $\expval{\hat{C}}{\bs{\gamma,\beta}}$ with respect to $(\bs{\gamma},\bs{\beta})$ using a good initial guess have been proposed in Ref.~\cite{Zhou_2020}. Let the approximation ratio (AR) be defined as
\begin{align}
  F_{p}(\bs{\gamma},\bs{\beta}) = \frac{\expval{H}{\bs{\gamma},\bs{\beta}}}{E_{\max}}.
\end{align}
We denote the variational parameters that optimize the expectation value by $(\bs{\gamma},\bs{\beta})_{AR}$. Rather than optimizing the expectation value, one may optimize the overlap of the ansatz state with the ground state $\ket{\psi_{\mathrm{GS}}}$,
\begin{align}
    \label{eq:TTS}
P_{p}(\bs{\gamma},\bs{\beta})
=
\left|
\left\langle
\psi_{\mathrm{GS}}
\middle|
\psi\!\left(
\boldsymbol{\gamma},
\boldsymbol{\beta}
\right)
\right\rangle
\right|^2.
\end{align}
We denote the parameters that optimize the ground-state overlap by $(\bs{\gamma},\bs{\beta})_{\mathrm{TTS}}$.

\subsection{AMP}

AMP algorithms are a class of local algorithms originally proposed in Ref.~\cite{AMP}. In spin-glass theory, AMP algorithms are closely related to the Thouless--Anderson--Palmer free-energy functional \cite{TAP,zou2022}. Assuming the absence of the OGP, there exists an AMP algorithm that approaches the optimal value of the Ising mean-field $k$-spin glass arbitrarily closely \cite{AMS21}. While specific AMP-type algorithms are known for the $k$-spin glass rather than for general Max-CSPs such as Max-$q$-XORSAT, it has been suggested that analogous AMP-type methods may exist more broadly \cite{JMSS23}. A survey of AMP and its variants can be found in Ref.~\cite{AMP_survey}. We introduce the finite-memory AMP class used in Ref.~\cite{rossetti2024linear, GJ21}. For fixed $M,T\in\mathbb{Z}^{+}$, consider functions $f_t:[-M,M]^t\rightarrow\mathbb{R}$ and $F_t:\mathbb{R}\times[-M,M]^t\rightarrow\mathbb{R}$ for $1\leq t\leq T$.

\begin{asm}
    \label{asm:AMP}
    $f_t(0)=0$. In addition, $f_t$ and $F_t$ are Lipschitz continuous on their respective domains. Specifically, there exists $K>0$ such that, for all $1\leq t\leq T$,
    \begin{align}
        |f_t(u)-f_t(v)|\leq K\|u-v\|_2\quad\text{for all }u,v\in[-M,M]^t,
        \\
        |F_t(u)-F_t(v)|\leq K\|u-v\|_2\quad\text{for all }u,v\in\mathbb{R}\times[-M,M]^t.
    \end{align}
\end{asm}

Fix $M>1$ and let $[x]_M=\max(-M,\min(x,M))$ denote coordinatewise truncation. Given $U^0\in[-M,M]^N$, define $U^t\in[-M,M]^N$ by
\begin{align}
\label{eq:Ut}
    U^t = \left[
        F_t\!\left(J(\cdot,f_t(U^0,\dots,U^{t-1})),U^0,\dots,U^{t-1}\right)
    \right]_M.
\end{align}
Here $f_t$, $F_t$, and the truncation are applied coordinatewise. At iteration $t$, the vector $f_t(U^0,\ldots,U^{t-1})$ is first formed coordinatewise. The tensor contraction $J(\cdot,f_t(U^0,\ldots,U^{t-1}))$ is then evaluated and supplied, together with the iteration history, to $F_t$. Finally, the result is truncated coordinatewise to obtain $U^t$. To generate a solution in $\mathcal{B}^N$, we assume access to a
projection function $\Pi_N : \mathcal{H}^N \to \mathcal{B}^N$.

\begin{algorithm}[h!]
\caption{Finite-memory AMP algorithm}\label{alg:AMP}
\begin{algorithmic}
\Require $J,U^0,M,T,(f_t)_{t=1}^{T},(F_t)_{t=1}^{T},\Pi_N$
\For{$t=1,\ldots,T$}
    \State Compute $U^t$ using Eq.~(\ref{eq:Ut}).
\EndFor
\State Set $V=[U^T]_1\in\mathcal{H}^N$.
\State Compute $\bs{z}^{*}=\Pi_N(V)$.
\State \Return $\bs{z}^{*}$.
\end{algorithmic}
\end{algorithm}

\subsection{MF-AOA}
The MF-AOA \cite{MFAOA} is a quantum-inspired classical analogue of QAOA in which quantum-spin evolution is replaced by classical mean-field spin dynamics. We restrict to the conventional linear interpolation $s(t)=t/T\in[0,1]$ used in Refs.~\cite{MFAOA,XOR_QAOA}, yielding
\begin{align}
    H(s)=(1-s(t))\sum_{i=1}^{N}X_i+s(t)C.
\end{align}
For the pure $q$-spin glass and Max-$q$-XORSAT without an external field, the initial symmetric state has vanishing effective magnetization. A catalyst field is therefore introduced,
\begin{align}
    H_{\mathrm{ex}}=\sum_i\lambda_i s^2(t)(1-s(t)),
\end{align}
where the $\lambda_i$ are sampled from the normal distribution $\mathcal{N}(0,\sigma^2)$ with mean $0$ and standard deviation $\sigma$. The effective magnetization of spin $i$ is
\begin{align}
    \label{eq:eff_mag}
    m_i(t)=\lambda_i s^2(t)(1-s(t))+
    \sum_{1\leq i_1,\ldots,i_{q-1}\leq N}
    J_{i,i_1,\ldots,i_{q-1}}
    n_{i_1}^{z}(t)\cdots n_{i_{q-1}}^{z}(t).
\end{align}
The spin vector $\bs{n}_i(p)$, with starting point $\bs{n}_i(0)=\bs{e}_x=(1,0,0)^{\mathsf T}$, evolves according to
\begin{align}
    \label{eq:MF_evolution}
    \bs{n}_i(p)=\prod_{t=1}^{p}V_i^D(t)V_i^P(t)\bs{n}_i(0),
\end{align}
with
\begin{subequations}
\label{eq:evolutions}
\begin{align}
V_i^D(t)&=
\begin{pmatrix}
1&0&0\\
0&\cos\!\bigl(2(1-s(t))\bigr)&\sin\!\bigl(2(1-s(t))\bigr)\\
0&-\sin\!\bigl(2(1-s(t))\bigr)&\cos\!\bigl(2(1-s(t))\bigr)
\end{pmatrix},\\
V_i^P(t)&=
\begin{pmatrix}
\cos\!\bigl(2m_i(t-1)s(t)\bigr)&\sin\!\bigl(2m_i(t-1)s(t)\bigr)&0\\
-\sin\!\bigl(2m_i(t-1)s(t)\bigr)&\cos\!\bigl(2m_i(t-1)s(t)\bigr)&0\\
0&0&1
\end{pmatrix}.
\end{align}
\end{subequations}
After the final iteration, the conventional MF-AOA returns $\bs{z}^{*}=\operatorname{sign}(\bs{n}^{z})$.

\subsection{Algorithms and the OGP}

\begin{dfn}[Overlap Gap Property \cite{OGP_Survey}]
For a maximization problem, the OGP holds if there exist $\epsilon>0$ and $0\leq\mu_1<\mu_2$ such that every pair of $\epsilon$-optimal configurations $\bs{z}^{1},\bs{z}^{2}\in\{-1,1\}^{N}$ satisfies
\begin{align}
H_J(\bs{z}^{a})\geq\max_{\bs{z}\in\{-1,1\}^{N}}H_J(\bs{z})-\epsilon,
\qquad a\in\{1,2\},
\end{align}
and its normalized overlap
\begin{align}
R_{1,2}=\frac{1}{N}\sum_{i=1}^{N}z_i^1z_i^2
\end{align}
satisfies
\begin{align}
|R_{1,2}|\in[0,\mu_1]\cup[\mu_2,1].
\end{align}
\end{dfn}

Given the OGP in the mean-field $q$-spin glass, Gamarnik and Jagannath established an obstruction for the finite-memory AMP class, conditional on an extension of the OGP from the binary cube $\mathcal{B}^N$ to the Hilbert cube $\mathcal{H}^N$ \cite{GJ21}.

\begin{asm}[Informal version of Conjecture~3.2 in Ref.~\cite{GJ21}]
\label{asm:OGP}
For every even $q\geq4$, the mean-field $q$-spin glass exhibits the OGP on $\mathcal{H}^N$ with probability at least $1-\exp(-cN)$ for sufficiently large $N$ and for some constant $c > 0$.
\end{asm}

\begin{thm}[Informal version of Theorem~3.3 in Ref.~\cite{GJ21}]
\label{thm:AMP_OGP}
For every even $q\geq4$, conditional on assumptions \ref{asm:AMP} and \ref{asm:OGP}, a fixed-step finite-memory AMP algorithm remains separated from the OGP energy level with probability at least $1-\exp(-cN)$ and for some constant $c > 0$.
\end{thm}
The theorem controls the AMP algorithm's output in $\mathcal{H}^N$ before an arbitrary final projection to a binary string.
\section{MF-AOA Within the Finite-Memory AMP Framework}
\label{sec:theory_arg}
We now show that the MF-AOA effective-field dynamics admits a natural finite-memory AMP representation. The purpose is to establish the structural correspondence required to invoke the obstruction result of Ref.~\cite{GJ21}. We consider a fixed number of iterations and, when applying the formal AMP theorem, the same coordinatewise truncation as in Eq.~(\ref{eq:Ut}). Let
\begin{align}
R_t(x)=V^D(t)V^P(t,x)
\end{align}
denote the one-step spin update when the effective magnetization is $x$. Starting from $\bs{n}_i(0)=\bs{e}_x=(1,0,0)^{\mathsf T}$, the $z$ component after $t$ iterations is a deterministic function of the local magnetization history. Define
\begin{align}
\phi_t(x_0,\ldots,x_{t-1})
=
\bs{e}_z^{\mathsf T}
R_t(x_{t-1})\cdots R_1(x_0)\bs{e}_x,
\label{eq:phi_history}
\end{align}
so that
\begin{align}
n_{i}^{z}(t)=\phi_t\bigl(m_i(0),\ldots,m_i(t-1)\bigr).
\label{eq:nz_history}
\end{align}
Although $V_i^P(1,0)=\mathbb{1}$, the driver rotation $V_i^D(1)$ need not be the identity; it merely leaves the initial vector $\bs{e}_x$ invariant. Set
\begin{align}
U_i^0=\lambda_i,
\qquad
U_i^{t+1}=m_i(t),
\qquad t\geq0.
\label{eq:AMP_MFAOA_iterates}
\end{align}
For $t\geq0$, define the coordinatewise functions
\begin{align}
f_{t+1}(u^0,\ldots,u^t)
&=\phi_t(u^1,\ldots,u^t),
\label{eq:ft_corrected}\\
F_{t+1}(h,u^0,\ldots,u^t)
&=h+u^0s^2(t)(1-s(t)).
\label{eq:Ft_corrected}
\end{align}
By construction,
\begin{align}
f_{t+1}(U_i^0,\ldots,U_i^t)=n_i^z(t).
\end{align}
Consequently,
\begin{align}
J\!\left(\cdot,f_{t+1}(U^0,\ldots,U^t)\right)_i
=
\sum_{i_1,\ldots,i_{q-1}}
J_{i,i_1,\ldots,i_{q-1}}
\prod_{a=1}^{q-1}n_{i_a}^{z}(t),
\label{eq:J_generates_field}
\end{align}
which is precisely the interaction contribution to Eq.~(\ref{eq:eff_mag}). Therefore,
\begin{align}
U_i^{t+1}
&=
F_{t+1}\!\left(
J\!\left(\cdot,f_{t+1}(U^0,\ldots,U^t)\right)_i,
U_i^0,\ldots,U_i^t
\right)
\notag\\
&=m_i(t).
\label{eq:MFAOA_AMP_correspondence}
\end{align}
With the coordinatewise truncation $[\cdot]_M$ included, Eq.~(\ref{eq:MFAOA_AMP_correspondence}) has exactly the finite-memory form of Eq.~(\ref{eq:Ut}).

\paragraph{Lipschitz continuous.}
We now verify that the continuous assumption~\ref{asm:AMP} holds. Since $V^D(t)$ and $V^P(t,x)$ are orthogonal,
\begin{align}
\|R_t(x)\|_{\mathrm{op}}=1.
\end{align}
The field dependence enters $V^P(t,x)$ through the angle $2s(t)x$. Hence
\begin{align}
\|R_t(x)-R_t(y)\|_{\mathrm{op}}
&\leq\|V^P(t,x)-V^P(t,y)\|_{\mathrm{op}}
\notag\\
&=2\left|\sin\bigl(s(t)(x-y)\bigr)\right|
\notag\\
&\leq2s(t)|x-y|.
\label{eq:rotation_lipschitz}
\end{align}
For $\bs{x}=(x_0,\ldots,x_{t-1})$ and $\bs{y}=(y_0,\ldots,y_{t-1})$, an expansion of the products in Eq.~(\ref{eq:phi_history}), together with the unit operator norm of all rotation matrices, gives
\begin{align}
|\phi_t(\bs{x})-\phi_t(\bs{y})|
&\leq
\sum_{r=1}^{t}\|R_r(x_{r-1})-R_r(y_{r-1})\|_{\mathrm{op}}
\notag\\
&\leq2\sum_{r=1}^{t}s(r)|x_{r-1}-y_{r-1}|
\displaybreak[3]
\notag\\
&\leq2\left(\sum_{r=1}^{t}s^2(r)\right)^{1/2}
\|\bs{x}-\bs{y}\|_2
\notag\\
&\leq2\sqrt{T}\,\|\bs{x}-\bs{y}\|_2,
\label{eq:phi_lipschitz}
\end{align}
for every fixed $t\leq T$. Thus $f_t$ is Lipschitz with a constant independent of $N$ for fixed $T$.

If the magnetization is zero, then $V^P(r,0)=\mathbb{1}$, while the driver rotations leave $\bs{e}_x$ invariant. Therefore
\begin{align}
f_t(0,\ldots,0)=\bs{e}_z^{\mathsf T}\bs{e}_x=0.
\end{align}
Finally, $F_{t+1}$ is affine in its first two relevant arguments. For two inputs $a=(h,u^0,\ldots)$ and $b=(h',v^0,\ldots)$,
\begin{align}
|F_{t+1}(a)-F_{t+1}(b)|
&\leq|h-h'|+s^2(t)(1-s(t))|u^0-v^0|
\notag\\
&\leq
\sqrt{1+\left(\frac{4}{27}\right)^2}\,\|a-b\|_2,
\end{align}
where $\max_{s\in[0,1]}s^2(1-s)=4/27$. Hence $F_t$ is also Lipschitz. We therefore conclude that, for fixed $T$ and with the truncation used in the finite-memory AMP definition, the MF-AOA satisfies the assumption~\ref{asm:AMP}. Conditional on the relaxed OGP assumption of Ref.~\cite{GJ21}, MF-AOA is unable to give a solution larger than the OGP value.
 
 \section{Numerical Analysis of QAOA}
  \label{sec:num_arg} 
By relating the MF-AOA to AMP algorithms in the previous section, we know that the QAOA will outperform the MF-AOA given sufficiently large circuit depth. It is thus an open question as to the depth required by the QAOA such that its expectation value is larger than the OGP threshold. Since no known analytic techniques exist to study the QAOA beyond logarithmic-depth, we perform a numerical analysis of the QAOA on instances of Max-$4$-XORSAT that exhibit the OGP and identify, for fixed $N$, the depth $p$ that is required to surpass the OGP threshold. As a comparison, we also evaluate the performance of the QAOA on weighted MaxCut (equiv. Max-$2$-XORSAT) instances which are believed to not exhibit the OGP \cite{CGPR19,JMSS23}.\\
Given an underlying hypergraph $G=(V,E)$ with $|V|=N$ vertices, and $|E|=m$ edges, the cost function for Max-$q$-XORSAT on such a hypergraph can be expressed as
\begin{align}
     H_{XOR}^q(\boldsymbol{z})= \sum_{(i_1,\dots,i_q)\in E} \frac{1}{2} (1+ J_{i_1 i_2 \dots i_q}z_{i_1} z_{i_2}\dots z_{i_q}),
\end{align}
with $\boldsymbol{z}\in \mathcal{B}^N$, and $ J_{i_1 i_2 \dots i_q}\in \{-1,1\}$.\\

For the instances of Max-$q$-XORSAT, we generate $d$-regular, $q$-uniform hypergraphs with the constraint that $Nd$ is a multiple of $q$. We also randomly generate the list $\bs{J}=\{-1,+1\}^{|E|}$ for the coupling strength of the hyperedges.\\

In order to identify where the OGP threshold is in each instance of Max-$4$-XORSAT, we perform a branch and bound algorithm and record those whose cut-fraction exceeds a certain threshold. From the set of solutions, we calculate the spectrum of the overlap and check if it is dense (i.e.\ all $N$ possible values in the interval $[0,1]$ are accounted for). If it is not dense, we lower the threshold and repeat until it is dense. Once we have the list of bit-strings and their corresponding cut-fraction, we set $\epsilon >0$ (typically $\epsilon=0.2$) such that the set of overlaps between $\epsilon$-optimal solutions is dense. We then decrease $\epsilon$ and repeat.
 If, for some $\epsilon$, there is a single gap in the spectrum, we use the instance generated for a case with the OGP. If no such $\epsilon$ exists, we use it as an instance without the OGP.
 
 \subsection{Average Performance: OGP vs No-OGP Instances}
\begin{figure}[h!]
    \centering
        \includegraphics[width=\textwidth]{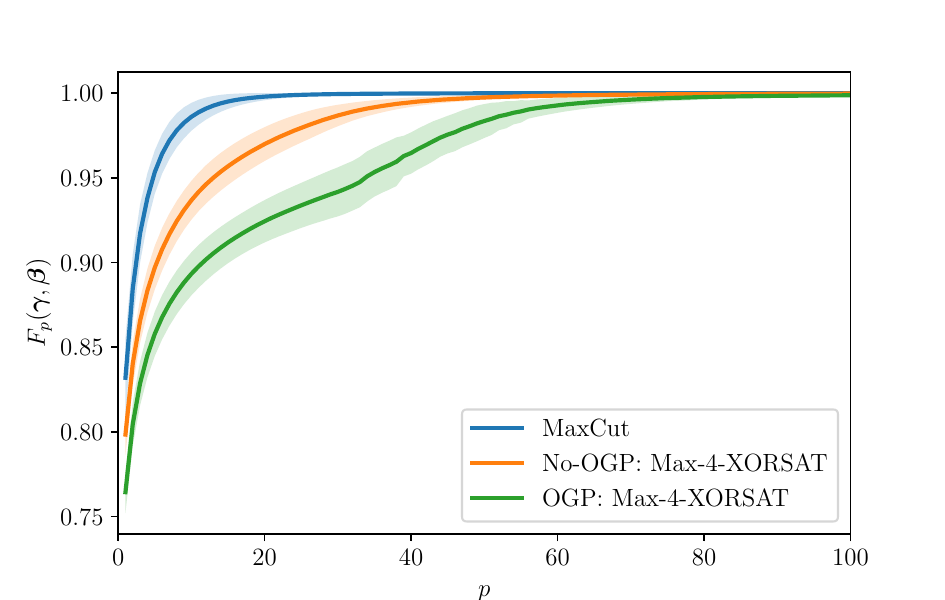}
    \caption{Mean QAOA approximation ratio ${F}_{p}$ as a function of depth $p$ and $N=15$. The blue curve corresponds to MaxCut instances. The orange curve represents Max-4-XORSAT instances without OGP barrier. The green curve shows Max-4-XORSAT instances exhibiting an OGP.
    }
    \label{fig:SE}
\end{figure}
We begin by comparing the QAOA performance averaged over multiple instances for the respective problem class. This provides a baseline for assessing how the emergence of OGP affects the depth dependence of the QAOA before examining the behavior of the QAOA on individual OGP instances in the following subsection. Fig.~\ref{fig:SE} compares MaxCut, Max-4-XORSAT instances without an OGP, and Max-4-XORSAT instances exhibiting an OGP. For MaxCut instances without clustering or overlap-gap structure, the approximation ratio rapidly saturates at shallow-depth, indicating that local quantum updates are sufficient to efficiently concentrate probability amplitude near optimal solutions. As compared to MaxCut, Max-4-XORSAT instances without OGP exhibit weak clustering, a solution-space geometry in which the majority of solutions can be partitioned into well-separated clusters, while a small exceptional set may still connect these clusters. In contrast to the OGP, such clustering therefore does not require a strictly forbidden interval of overlaps \cite{OGP_Survey}. In contrast, Max-4-XORSAT instances without the OGP already exhibit noticeably slower convergence, suggesting that weak clustering effects partially obstruct the redistribution of probability amplitudes across configuration space. The strongest slowdown occurs for OGP-Max-4-XORSAT instances, where the fragmented geometry of near-optimal solutions is associated with the need to redistribute amplitude across widely separated clusters. In this regime, shallow local updates become inefficient, consistent with recent theoretical results showing limitations of local quantum algorithms in overlap-gap structures \cite{chou_et_al:LIPIcs.ICALP.2022.41}. The clear separation between the three curves therefore provides numerical evidence that the geometric structure of the solution space strongly influences the depth scaling required for QAOA to reach near-optimal states.
 \subsection{Non-Differentiable Point at Critical Depth $p^*$}
 \label{susbec:42}
\begin{figure}[t]
    \centering
    \includegraphics[width=\linewidth]{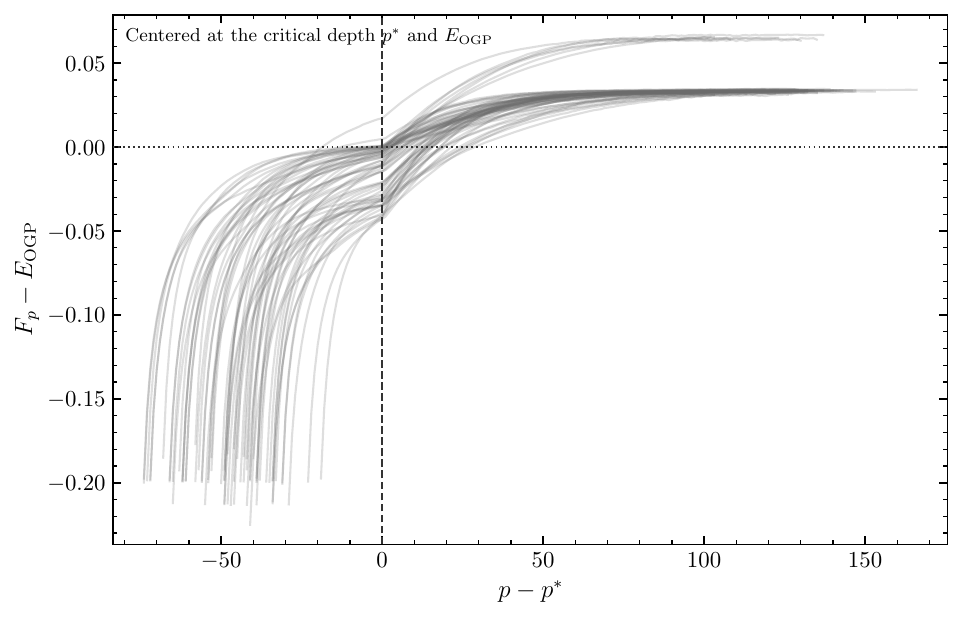}
    \caption{
    The data as shown in Fig.~\ref{fig:AppendixEnergyOriginal} after translating every $F_{p}$ of each instance according to $p-p^{*}$ and $F_p-E_{\mathrm{OGP}}$. The vertical dashed line indicates the instance-dependent critical depth $p^{*}$, while the horizontal dotted line denotes the energy at OGP emergence.
    }
    \label{fig:AppendixEnergyShifted2}
\end{figure}

\begin{figure}[t]
\centering
    \includegraphics[width=\linewidth]{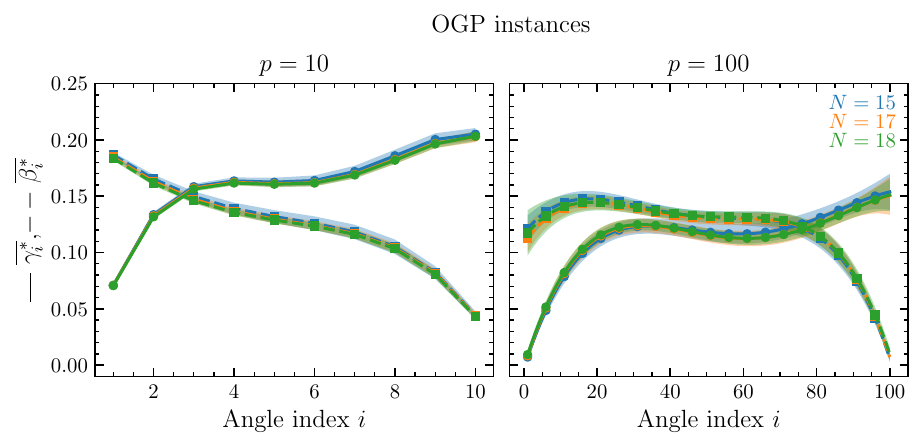}
    \caption{OGP instances:
    Average optimized QAOA parameter schedules obtained using the Fourier parameterization of Ref.~\cite{Zhou_2020}. Solid and dashed curves denote the averaged phase-separation angles $\overline{\gamma_i^{*}}$ and mixer angles $\overline{\beta_i^{*}}$, respectively, while different colours correspond to different system sizes $N$.%
    }
    \label{fig:AppendixParameters}
\end{figure}

While the previous analysis focused on the ensemble averages, analyzing individual problem realizations reveals additional structure in the optimization dynamics of the QAOA.
 \begin{figure}[h!]
    \centering
        \includegraphics[width=\textwidth]{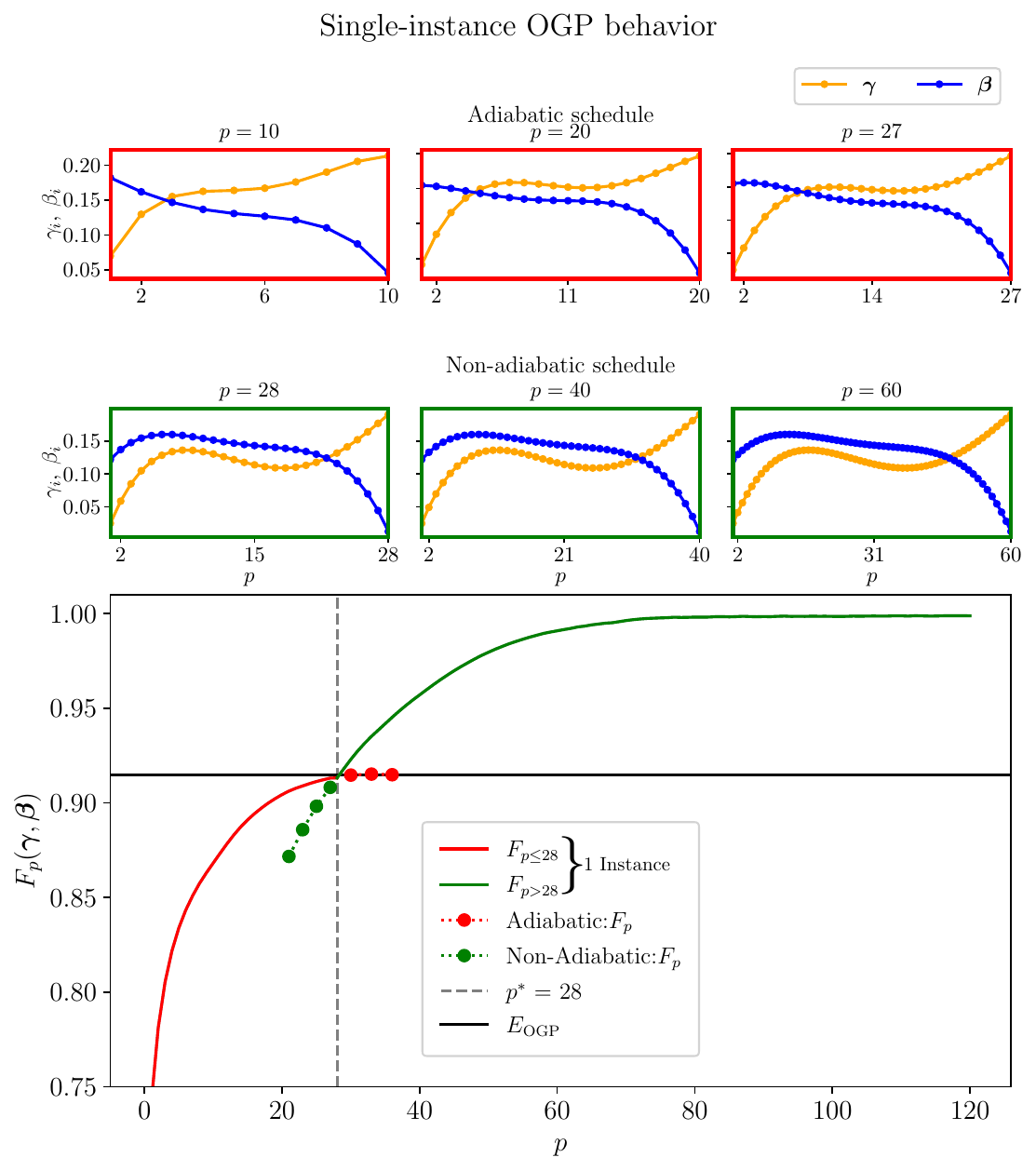}
    \caption{QAOA approximation ratio $F_{p}$ as a function of circuit depth $p$ for a single N=15 Max-4-XORSAT instance exhibiting a transition near the critical depth $p^{*} = 28$ (gray dashed line). The red curve ($F_{p \leq 28}$) corresponds to optimization trajectories obtained from an approximately adiabatic parameter schedule (first row, red frames), while the green curve ($F_{p > 28}$) corresponds to a non-adiabatic parameter branch (second row, green frames) emerging beyond the critical depth. Red and green markers indicate the achieved approximation ratios when the respective adiabatic and non-adiabatic schedules are explicitly followed, rather than fully re-optimized at each depth. The inset panels show the associated optimized QAOA parameters $\boldsymbol{\gamma},\boldsymbol{\beta}$ for the two regimes.}
    \label{fig:Single_Instance_OGP}
\end{figure}
For instances that exhibit the OGP, we observe the emergence of a non-differentiable point at a critical depth $p^*$, where the scaling behavior of the QAOA expectation value $\expval{H}{\bs{\gamma},\bs{\beta}}$ changes. Fig.~\ref{fig:Single_Instance_OGP} shows a representative $N=15$ Max-4-XORSAT instance exhibiting a critical depth $p^{*} \approx 28$, where the optimized QAOA parameters undergo a transition between two distinct parameter regimes. For shallow-depth $p<p^{*}$, the optimal QAOA parameters follow an approximately linear schedule in $p$, characteristic of an adiabatic-like evolution, as illustrated in the first row (red frames) of Fig.~\ref{fig:Single_Instance_OGP}. In this regime, the approximation ratio improves only gradually and saturates close to the OGP threshold. Near the critical depth $p^{*}$, the optimization landscape reorganizes and the adiabatic parameter branch ceases to yield improved performance. Instead, the globally optimal parameters transition into a non-adiabatic schedule, shown in the second row (green frames) of Fig.~\ref{fig:Single_Instance_OGP}. Once this non-adiabatic branch is reached, the QAOA performance rapidly improves and approaches the ground-state energy at larger depth. We observe this behavior for the majority of OGP instances studied, suggesting that the OGP induces not only a geometric obstruction in configuration space, but also a restructuring of the variational optimization landscape itself as the circuit depth $p$ increases. In Fig.~\ref{fig:Single_Instance_OGP}, the green and red dots indicate that extrapolating the respective parameter schedules without full re-optimization is insufficient to recover the globally optimal solution, demonstrating that the transition is not a smooth continuation of the shallow-depth adiabatic trajectory. 
Since the critical depth $p^{*}$ is instance dependent, the non-differentiable point is averaged out in the ensemble results shown in Fig.~\ref{fig:SE}. To demonstrate that this behavior is a generic feature of the OGP instances, Fig.~\ref{fig:AppendixEnergyShifted2} shows the QAOA performance of all individual instances after aligning the trajectories with respect to their critical depths $p^*$ along the horizontal axis and their respective OGP energies $E_{\mathrm{OGP}}$ along the vertical axis. The aligned trajectories reveal that the non-differentiable point occurs in the vicinity of the OGP energy, typically slightly below the OGP threshold, demonstrating that the behavior observed for the representative instance in Fig.~\ref{fig:Single_Instance_OGP} is a generic feature of the investigated OGP instances. The corresponding unaligned trajectories are shown in Appendix~A,
Fig.~\ref{fig:AppendixEnergyOriginal}.\\
Fig.~\ref{fig:AppendixParameters} shows the averaged optimized QAOA parameters, showing that the change in the optimal parameter schedules observed in Fig.~\ref{fig:Single_Instance_OGP} is not specific to a single instance, but persists across OGP instances of different system sizes.
To avoid the rapidly changing non-differentiable point region near the instance-dependent critical depth $p^{*}$, the averages are computed at two representative circuit depths well below ($p=10$) and well above ($p=100$) the transition. The resulting schedules exhibit universality and provide further numerical evidence for the transition from an approximately adiabatic to a non-adiabatic parameter regime.\\
\begin{figure}[h!]
\centering
    \includegraphics[width=\linewidth]{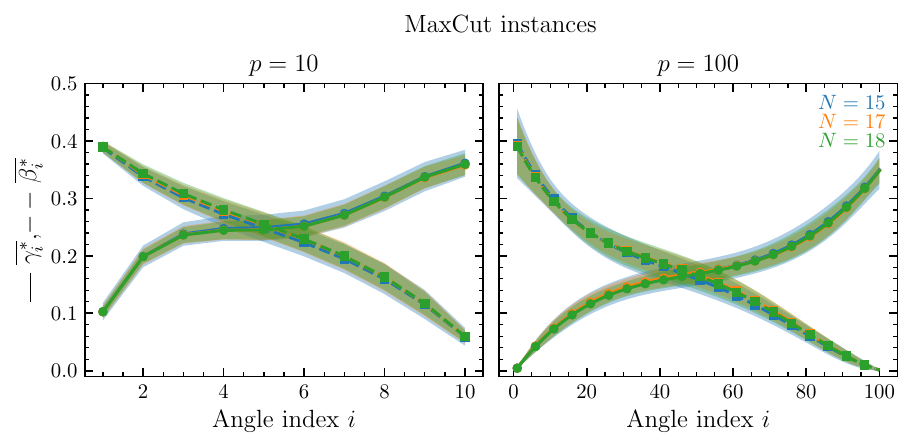}
    \caption{MaxCut instances: For comparison with the OGP instances in Fig.~\ref{fig:AppendixParameters},
we show the average optimized QAOA parameter schedules for MaxCut at $p=10$ and $p=100$ for different system sizes. As for the OGP instances, the parameters are obtained using the Fourier parameterization of Ref.~\cite{Zhou_2020} and averaged over the considered instances. Solid and dashed curves denote the averaged phase-separation angles $\overline{\gamma_i^{*}}$ and mixer angles $\overline{\beta_i^{*}}$, respectively.
}
    \label{fig:AppendixParameters2}
\end{figure}
Fig.~\ref{fig:AppendixParameters2} provides the corresponding averaged optimized parameter schedules for MaxCut instances without OGP structure. In contrast to the OGP instances, the schedules retain a smooth
adiabatic-like form with increasing depth, with no indication of the
distinct non-adiabatic parameter structure observed for OGP instances.
The close agreement across different system sizes further demonstrates
that this behavior is not instance- or size-specific, consistent with the concentration of the optimized MaxCut schedules already observed for the AR-optimized QAOA parameters.
\begin{figure}[h!]
    \centering
        \includegraphics[width=\textwidth]{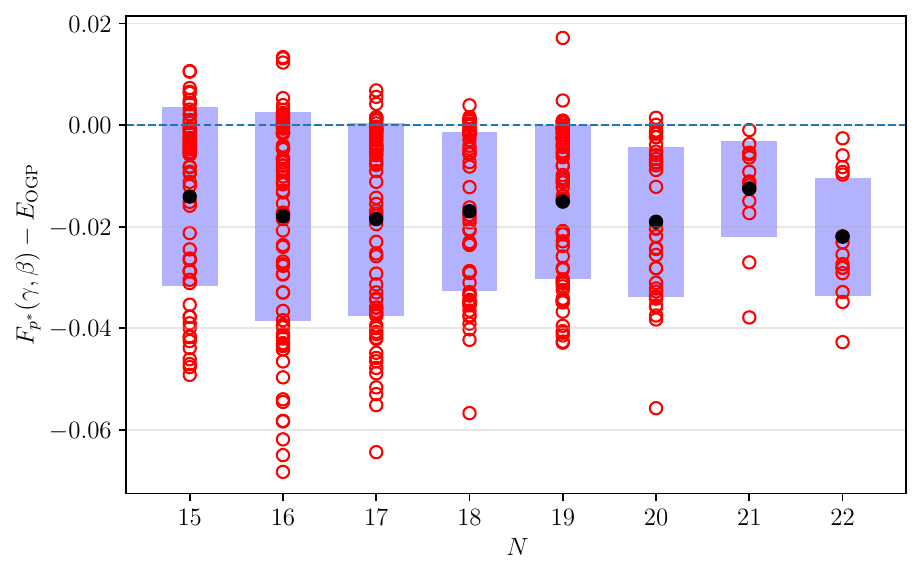}
    \caption{Difference between the critical transition depth $p^{*}$
 and the OGP onset for all studied Max-4-XORSAT instances as a function of system size $N$.$F_{p^*}(\gamma,\beta) - E_{\mathrm{OGP}}$ where $E_{\mathrm{OGP}}$ is the instance-specific energy level where the OGP occurs. Each red marker corresponds to a result for a single instance, while the semi-transparent blue regions indicate the standard deviation of the distribution for each system size and black markers correspond to the mean for each system size. Negative values indicate that the transition-like restructuring of the QAOA optimization dynamics occurs before or near the onset of the overlap-gap regime.}
    \label{fig:Sample_Instances}
\end{figure}
To investigate the relationship between the critical depth $p^{*}$ and the emergence of overlap-gap structure, Fig.~\ref{fig:Sample_Instances} shows  the difference between $p^{*}$ and the OGP onset for all studied instances  as a function of system size $N$. The majority of instances exhibit $p^*-\mathrm{OGP}<0$, indicating that the transition from adiabatic-like to non-adiabatic parameter schedules typically occurs slightly before the QAOA expectation value reaches the OGP threshold. Nevertheless, the small magnitude of this offset demonstrates that the two phenomena occur in close proximity. This behavior remains consistent across all investigated system sizes $N$ and suggests a close connection between the OGP barrier and the breakdown of the shallow-depth adiabatic optimization regime observed previously. The relatively narrow spread of the distributions further indicates that this phenomenon is not dominated by rare disorder realizations, but instead represents a robust characteristic of OGP instances. Additionally, the variance of the spread also decreases as system size $N$ increases. Together with the single-instance analysis across large $p$, these results support the interpretation that clustered solution-space geometry is associated with a restructuring of the variational optimization landscape, requiring the QAOA to transition into a distinct non-adiabatic regime in order to continue improving performance beyond the OGP threshold. The emergence of the non-differentiable point depends on the objective used to optimize the QAOA parameters. In Subsec.~\ref{subsec:43}, we briefly demonstrate that optimizing the ground-state overlap, rather than the expectation value, removes the non-differentiable point for the instances considered there. This indicates that the optimal variational parameters $(\boldsymbol{\gamma}, \boldsymbol{\beta})$ can depend qualitatively on the chosen optimization objective. We leave a systematic investigation of this dependence, and its potential use for heuristic strategies to select the variational parameters,
to future work.\\
To verify that the observed change in the scaling of QAOA for OGP instances is not an artifact of the classical optimization procedure, we performed several numerical consistency checks. First, we repeated the parameter optimization using different classical optimizers, including  COBYLA, BFGS, and Nelder--Mead, and consistently recovered the same qualitative change in the optimal parameter schedules near $p^{*}$. In addition, the optimization was repeated over many initializations, combining random initial parameters with the Fourier-based parameterization, to reduce the possibility that the observed transition originates from a particular local optimum. Finally, applying the same optimization procedure to MaxCut instances of comparable system size does not produce a corresponding non-differentiable point in the QAOA approximation ratio, while the optimized parameters retain the smooth, adiabatic-like structure shown in Fig.~\ref{fig:SE}. Taken together, these checks strongly suggest that the transition is a property of the OGP instances rather than an artifact of the optimizer, initialization strategy, or finite parameterization.
\subsection{Non-Polynomial Scaling Behavior in OGP Instances}
\label{subsec:43}
\begin{figure}[H]
    \centering
    \includegraphics[
        width=\linewidth,
        height=0.48\textheight,
        keepaspectratio
    ]{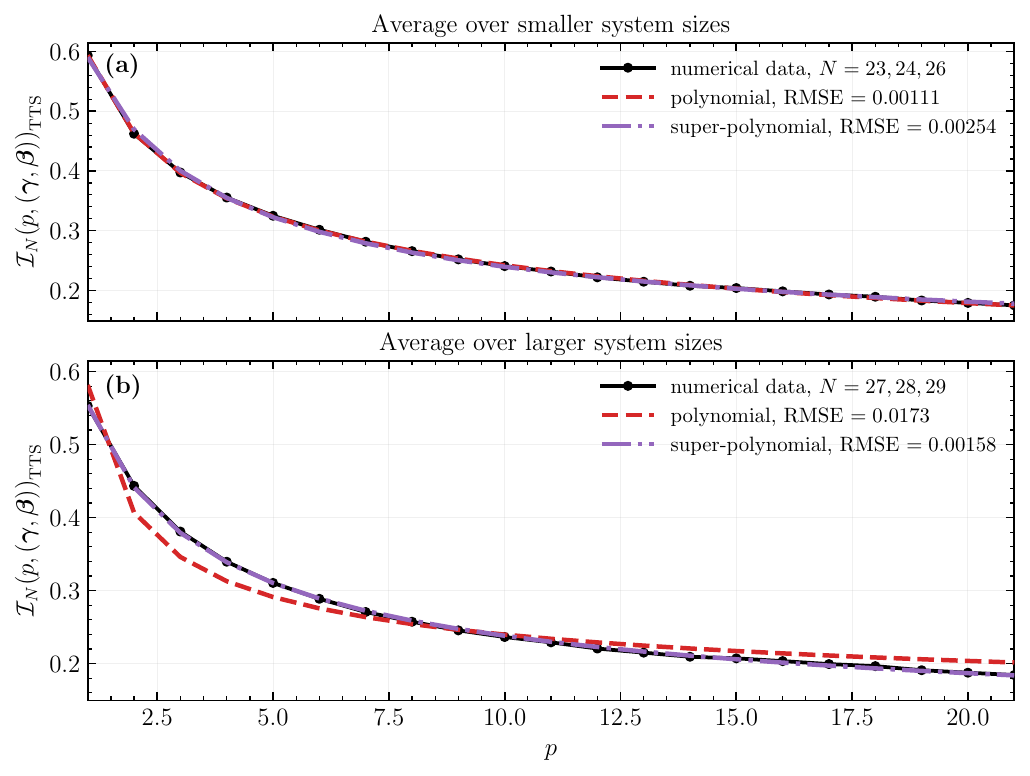}
   \caption{Panels (a) and (b) show the decay rate defined in Eq.~(\ref{eq:InverseOverlapRate}). In panel (a), the transformed curves for the three smaller system sizes are averaged pointwise over their common depth range; panel (b) shows the corresponding pointwise average for the three larger system sizes. Black lines with markers denote the averaged numerical data. The red dashed and purple dash-dotted curves show, respectively, polynomial fits and super-polynomial fits defined in Eq.~(\ref{eq:cases}).}
   \label{fig:Scaling_quasi_vs_poly}
\end{figure}

\begin{figure}[h!]
    \centering
        \includegraphics[width=\textwidth]{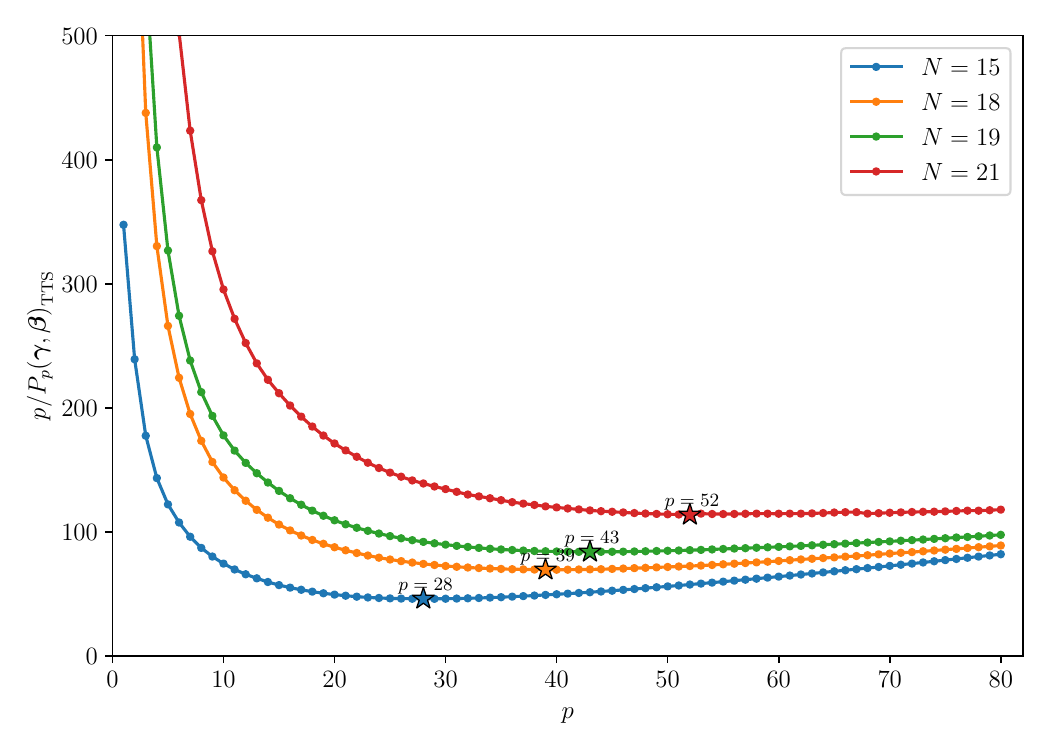}
    \caption{Time-to-solution (TTS) proxy for exact ground-state sampling as a function of the QAOA circuit depth $p$ for OGP Max-4-XORSAT instances of different system sizes. The plotted quantity, $p / P_{p}{(\boldsymbol{\gamma},
\boldsymbol{\beta})}_{\mathrm{TTS}}$, combines the linear increase in circuit cost with depth and the inverse probability of sampling the exact ground state. The star markers indicate the circuit depth $p$ minimizing the TTS proxy.
}
    \label{fig:Inv_scaling}
\end{figure}

Beyond finding the depth required of the QAOA to surpass the OGP barrier for fixed $N$, we also optimized the QAOA with respect to finding the ground state to identify the dependence of the scaling coefficient on $p$ via the equation
\begin{equation}
P_{p}{(\boldsymbol{\gamma},
\boldsymbol{\beta})}_{\mathrm{TTS}}
=
2^{-N\,\mathcal{I}_N(p,{(\boldsymbol{\gamma},
\boldsymbol{\beta})}_{\mathrm{TTS}})}
\label{eq:IR}
\end{equation}

\begin{equation}
\Leftrightarrow \mathcal{I}_N(p,{(\boldsymbol{\gamma},
\boldsymbol{\beta})}_{\mathrm{TTS}})
=
\frac{1}{N}
\log_{2}
\left(
\frac{1}{P_{p}{(\boldsymbol{\gamma},
\boldsymbol{\beta})}_{\mathrm{TTS}}}
\right),
\label{eq:InverseOverlapRate}
\end{equation}
with 
\begin{equation}
P_{p}{(\boldsymbol{\gamma},
\boldsymbol{\beta})}_{\mathrm{TTS}}
=
\left|
\psi_{\mathrm{GS}}
\middle|
\psi\!\left(
{\boldsymbol{\gamma},
\boldsymbol{\beta})}_{\mathrm{TTS}}
\right\rangle
\right|^2,
\end{equation}
being the  overlap with the ground state while optimizing for the ground state ($(\bs{\gamma},\bs{\beta})_{\mathrm{TTS}}$). As anticipated in Subsec.~\ref{susbec:42}, in contrast to the approximation-ratio
optimization, we do not observe a non-differentiable point when optimizing the ground-state overlap for the instances considered here. This allows us to perform a continuous scaling analysis. Whether the QAOA is able to find the optimal solution with polynomial circuit depth depends on the scaling exponent $c(p)$. Here, we perform two different numerical fits which correspond to the different time complexities often encountered in the analysis of algorithms
\begin{align}
\label{eq:cases}
    \mathcal{I}_N(p,{(\boldsymbol{\gamma},
\boldsymbol{\beta})}_{\mathrm{TTS}})=
    \begin{cases}
    \displaystyle
    \frac{c_1}{p^{c_2}+c_3} & \text{polynomial},\\
     \frac{c_1}{\log (p+c_3)^{c_2}} & \text{super-polynomial}
    \end{cases}
\end{align}
For Max-4-XORSAT instances exhibiting the OGP, our results provide evidence for a crossover away from polynomial-depth scaling as the system size increases, as shown in Fig.~\ref{fig:Scaling_quasi_vs_poly}. For the smaller system sizes [Fig.~\ref{fig:Scaling_quasi_vs_poly}(a)], the polynomial and super-polynomial fits are nearly indistinguishable, yielding comparable RMSE values of $\mathrm{RMSE}_{\mathrm{poly}}=1.11\times10^{-3}$ and $\mathrm{RMSE}_{\mathrm{s-poly}}=2.54\times10^{-3}$, respectively. In contrast, for the larger system sizes [Fig.~\ref{fig:Scaling_quasi_vs_poly}(b)], the polynomial fit develops a systematic deviation from the numerical data at increasing circuit depth, whereas the super-polynomial still fits. To quantify the relative quality of the two scaling descriptions beyond the visual comparison, we consider the ratio of their RMSE values. While the polynomial fit performs slightly better for the smaller system sizes, the super-polynomial fit reduces the RMSE by approximately one order of magnitude for the larger systems:
\begin{equation}
\left.
\frac{\mathrm{RMSE}_{\mathrm{poly}}}
     {\mathrm{RMSE}_{\mathrm{s-poly}}}
\right|_{N=23,24,26}
\simeq 0.44
\qquad\qquad
\left.
\frac{\mathrm{RMSE}_{\mathrm{poly}}}
     {\mathrm{RMSE}_{\mathrm{s-poly}}}
\right|_{N=27,28,29}
\simeq 10.95.
\end{equation}
The emerging preference for the super-polynomial description at larger \(N\) indicates that a polynomial increase in circuit depth is insufficient to overcome the OGP-induced limitation within the investigated regime. Instead, our results are consistent with the depth required to overcome this barrier exhibiting super-polynomial, and hence sub-exponential rather than polynomial, scaling.\\
Lastly, we investigate the circuit depth that minimizes the overall computational effort required to obtain the exact ground state. While increasing the QAOA depth improves the overlap with the ground state, deeper circuits also incur a proportionally larger implementation cost. Consequently, the relevant quantity is not the overlap alone but the time-to-solution (TTS), approximated by  
\begin{equation}
\mathrm{TTS}_{p} = \dfrac{p}{P_{p}{(\boldsymbol{\gamma},
\boldsymbol{\beta})}_{\mathrm{TTS}}},
\end{equation}
which combines the circuit depth with the inverse success probability of sampling the ground state. The resulting TTS proxy is shown in Fig.~\ref{fig:Inv_scaling} for several OGP Max-4-XORSAT system sizes. For all investigated instances, the curves exhibit a minimum at finite circuit depth, indicating an optimal trade-off between increasing the success probability and the additional cost of deeper circuits. Beyond this optimum, the ground-state overlap continues to improve only marginally, while the circuit cost grows linearly with p, resulting in diminishing returns and an increasing TTS proxy. 
Consequently, running the QAOA at the largest accessible depth is generally not optimal for exact ground-state sampling. An important observation is that the optimal depth $p$ increases only moderately with system size and remains substantially smaller than the depths that would be required to maximize the ground-state overlap itself. Therefore, even if achieving asymptotically large overlaps requires increasingly deep circuits, the depth minimizing the expected computational effort remains comparatively modest for the system sizes considered here. The optimal QAOA depth is determined by the balance between success probability and circuit cost rather than by maximizing the ground-state overlap alone.

\subsection{Passing the OGP Barrier for Increasing $N$}
In this subsection, we focus on the depth $p$ at which the QAOA first reaches the OGP barrier and interpret it as a meaningful algorithmic threshold rather than merely
an empirical marker. 
\begin{figure}[h!]
    \centering
        \includegraphics[width=\textwidth]{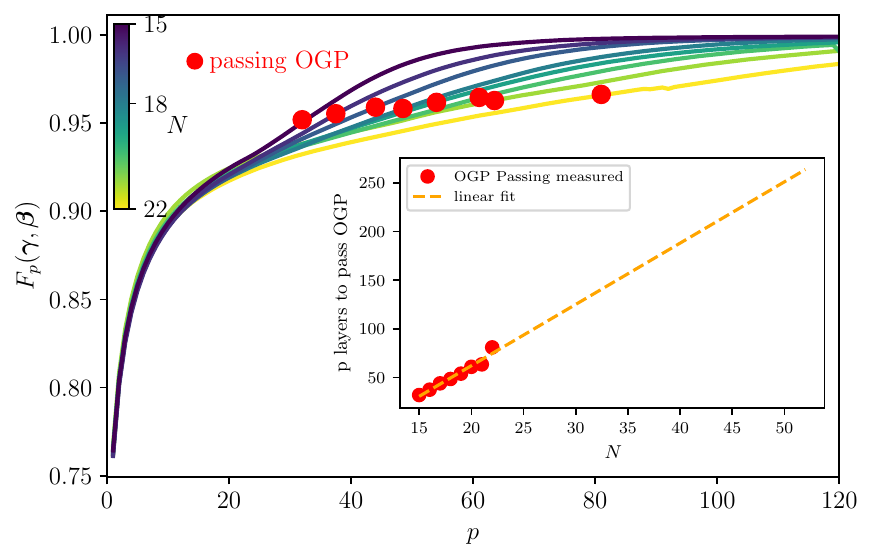}
    \caption{Mean energy curves for QAOA on sparse Max-4-XORSAT instances in the OGP regime.
Shown is the average approximation ratio as a function of QAOA depth $p$ for system sizes N=15,…,22 (viridis color scale). The approximation ratio is normalized such that values closer to unity correspond to energies closer to the ground-state energy. Red markers indicate the average depth $p$ at which the QAOA trajectory reaches the OGP barrier for the corresponding system size, averaged over disorder realizations. The inset replots these crossing points versus N and includes a linear fit as a best case extrapolation of the barrier-crossing depth.}
    \label{fig:OGP_Passing}
\end{figure}
Fig.~\ref{fig:OGP_Passing} provides finite-size evidence that the QAOA circuit depth required to overcome the OGP barrier increases with the problem size. Passing the OGP barrier is important because it separates two different algorithmic tasks. At and below the barrier, low-depth QAOA typically prepares states that are localized within one of several disconnected near-optimal clusters in overlap space. In this sense, crossing the OGP threshold serves as a proxy for whether the Ansatz can generate the long-range correlations required to effectively tunnel from a sub-optimal cluster into the cluster containing the optimal solution. To estimate the system size at which the QAOA may become advantageous over classical algorithms, we perform a linear extrapolation in Fig.~\ref{fig:OGP_Passing} up to $N=50$ qubits, in order to identify the regime where the OGP barrier may be crossed. We need roughly $p=250$ layers at $N=50$ to surpass the OGP barrier. We stress that the linear fit in Fig.~\ref{fig:OGP_Passing} represents an optimistic finite-size extrapolation and should not be interpreted as an asymptotic scaling law. Even under this favorable assumption, however, the required barrier-crossing depths grow rapidly with system size and, as shown in Subsec.~\ref{subsec:RE}, already impose resource requirements beyond those accessible to current quantum hardware. The regime $N=50$ represents a crossover point between classically tractable and computationally possible problem sizes. In classical simulation literature, exact Schrödinger simulation requires significant memory usage nearing physical limitations \cite{ circuit_simulation_hard}, and 50–60 qubits mark the onset at which universal circuit simulation becomes hard and, for highly entangled circuits, intractable. This is the scale for the system sizes targeted in first quantum advantage experiments, establishing it as the first nontrivial regime where quantum devices may outperform classical methods \cite{Arute2019}.
For sparse Max-4-XORSAT OGP instances, this scale is significant because exact branch-and-bound certification becomes exponentially demanding while local classical heuristics are expected to remain obstructed by the clustered near-optimal geometry \cite{CHM23}. Consequently, the vicinity of 50 qubits constitutes the first finite-size regime in which a possible QAOA crossing of the OGP barrier would become algorithmically and computationally meaningful.

\subsection{Resource Estimation to Cross the OGP Barrier} \label{subsec:RE}
 We combine numerical observations of OGP crossing with a resource-based analysis of QAOA circuits. Leveraging the locality of the Max-4-XORSAT Hamiltonians and the sparse structure of the underlying interaction graph, we derive explicit estimates for the number of entangling gates required to reach the OGP barrier. If the depth needed for crossing the OGP barrier grows too quickly, the circuit cost becomes incompatible with scalable implementation. Therefore, we present a short resource estimation while comparing with current hardware fidelities. For our analysis we use CNOT gates as our two-qubit entangling gate. For each clause in the Ansatz we need 3 CNOT gates for parity one single-qubit $Z$-rotation gate and 3 CNOTs for uncomputing.
\begin{figure}[h!]
    \centering
        \includegraphics[width=\textwidth]{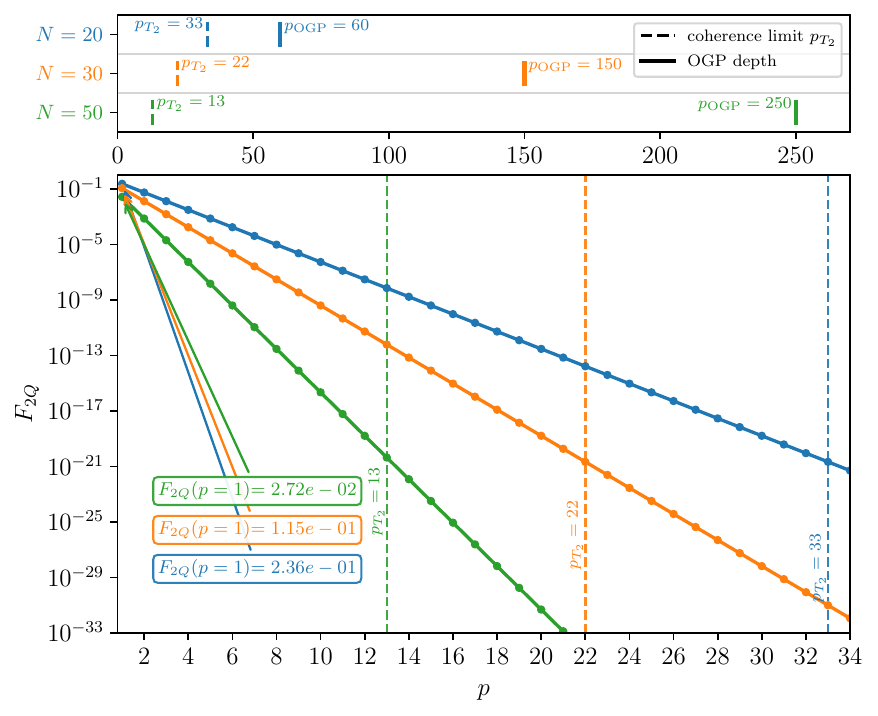}
    \caption{Top panel: Comparison between the coherence-limited depth $p_{T_{2}}$ (dashed) and the estimated OGP depth $p_{\text{OGP}}$
 (solid). Bottom panel: Estimated QAOA circuit fidelity as a function of depth $p$ for IBM superconducting hardware under the independent two-qubit error model of Eq.~(\ref{eq:circuit_fidelity}). Curves correspond to problem sizes $N=20,30,50$. The dashed vertical line indicates the coherence-limited depth $p_{T_2}$, while the boxes on the right show the remaining distance to the estimated OGP depth.}
    \label{fig:resource_estimation}
\end{figure}
To estimate the impact of hardware noise and decoherence time, we employ a simplified two-qubit gate error model and neglect single-qubit gate and SPAM errors. Since two-qubit operations dominate both the error budget and the circuit runtime on current superconducting processors, the total circuit fidelity is approximated as \cite{9251243}
\begin{equation}\label{eq:circuit_fidelity}
F_{2Q}=\prod_{i=1}^{G_t}(1-e_{2Q})
\end{equation}
where $e_{2Q}$ is the two-qubit gate error rate and $G_t$ is the total number of executed two-qubit gates. This model assumes statistically independent errors and therefore provides an upper bound estimation of the achievable circuit fidelity. To account for parallel execution, the effective circuit depth per QAOA layer is estimated as
\cite{Kaldenbach_2025}
\begin{equation}
D_{p} = \frac{G_{p}}{\Delta+1},
\end{equation}
where $G_{p}$ denotes the number of two-qubit gates per QAOA layer and $\Delta$ is the maximum graph degree \cite{Kaldenbach_2025}. The coherence-limited QAOA depth is then approximated by
\begin{equation}
p_{T_2} = \frac{T_2}{D_{p} \, t_{2Q}},
\end{equation}
with $t_{2Q}$ the two-qubit gate duration and $T_2$ the qubit coherence time. Fig.~\ref{fig:resource_estimation} Shows the analysis of decoherence and circuit fidelity with respect to superconducting hardware. The circuit fidelity decreases exponentially with QAOA depth and system size, reaching values far below practically useful accuracies well before the coherence-limited depth $p_{T_2}$ is attained. More importantly, the coherence-limited depth is reached long before the OGP regime becomes accessible. Even under the idealized assumption of perfect gate fidelity, the circuit runtime would exceed the available coherence time before reaching the predicted OGP depth. These results suggest that coherence limitations, in addition to gate errors, prevent current superconducting hardware from probing the QAOA depths associated with OGP-related algorithmic barriers. For the numerical evaluation we use representative IBM Heron hardware parameters, namely a two-qubit gate fidelity of approximately $99.7\%$, a two-qubit gate execution time of $200\,\mathrm{ns}$, and a coherence time of $300\,\mu\mathrm{s}$ \cite{ibm_quantum_processors,postquantum_ibm_heron_2025}.

\section{Discussion and Conclusion}
In this paper, we have established a connection between MF-AOA and finite-memory AMP, showing that MF-AOA is subject to the corresponding OGP-induced performance obstruction under the assumptions discussed in Sec.~\ref{sec:theory_arg}. As a consequence, one avenue for quantum advantage offered by the QAOA over its classical counterpart would be its ability to provide solutions that exceed this threshold. Our analysis indicates that a linear extrapolation of the observed finite-size trend yields an estimated critical circuit depth of approximately $p\approx250$ at $N=50$, where the QAOA first surpasses the OGP barrier. As $N=50$ is commonly regarded as approaching the limits of exact classical simulation \cite{circuit_simulation_hard}, this extrapolation provides a useful finite-size benchmark for the circuit depths that may be required in practically relevant regimes. We emphasize, however, that this linear extrapolation should not be interpreted as the asymptotic scaling of the critical depth. Also, by investigating the optimization with respect to the ground-state overlap for large system sizes, we find numerical evidence that the dependence on circuit depth is more accurately described by a super-polynomial rather than a polynomial scaling. It is interesting to note that the MF-AOA, as an AMP algorithm, is a general-purpose algorithm compared to those that are problem-specific such as the ones used to solve the mean-field $q$-spin glass. It would be noteworthy to see if the MF-AOA is comparable in performance or time complexity for such problems. On a related note, given that the optimal parameters of the QAOA follow a non-adiabatic schedule at larger depths, does optimizing the MF-AOA to follow a non-adiabatic schedule result in a better overall performance as well? Though we have proven that the MF-AOA will be unable to surpass the OGP barrier, it remains an open question as to how close the MF-AOA can get to the algorithmic threshold and whether this is best achieved via an adiabatic or non-adiabatic route.

\section*{Acknowledgement}
This project was made possible by the DLR Quantum Computing Initiative (QCI) and the Federal Ministry for Research, Technology and Space. M.G. was funded by the DLR QCI through the Quantum Fellowship Programme. M.G. thanks Kunal Marwaha for clarifying the results of logarithmic-depth limitations of the QAOA and David Gross for insightful discussions. The authors gratefully acknowledge the scientific support and HPC resources provided by the DLR. The HPC system CARO is partially funded by the Ministry of Science and Culture of Lower Saxony and the Federal Ministry for Research, Technology and Space.

\printbibliography 

\pagebreak 

\appendix

\renewcommand{\thesection}{Appendix \Alph{section}:}

\section{Analysis of the OGP Transition}
\label{app:ogp_transition}
\begin{figure}[h!]
    \centering
    \includegraphics[width=\linewidth]{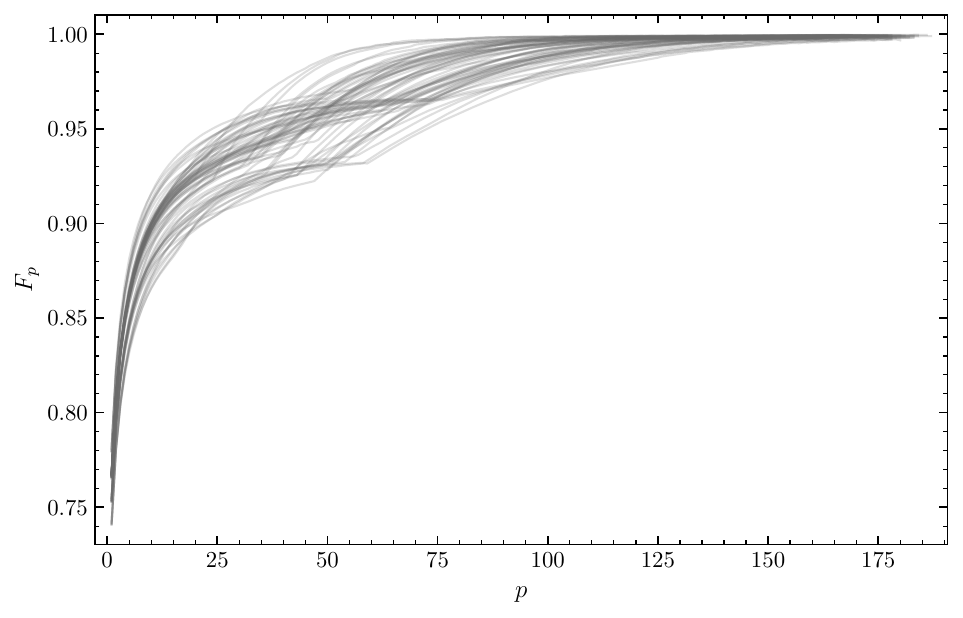}
    \caption{
    Evolution of the QAOA energy for all OGP instances as a function of circuit depth $p$ for $N=19$. Each grey curve corresponds to one instance of Max-4-XORSAT.
    }
    \label{fig:AppendixEnergyOriginal}
\end{figure}

In this appendix, we provide additional numerical results supporting the analysis of the overlap-gap property (OGP) presented in the main text. Fig.~\ref{fig:AppendixEnergyOriginal} shows the QAOA energy for all OGP instances as a function of circuit depth. At shallow circuit depths, all instances exhibit nearly identical optimization dynamics and therefore almost all $F_{p}$ overlap. As the circuit depth approaches the instance-dependent critical depth $p^*$, the trajectories separate as different disorder realizations undergo the transition at different depths. These critical depths occur in close proximity to, but do not generally coincide exactly with, the depths at which the corresponding QAOA trajectories reach the OGP threshold.
To expose the underlying universal behavior, every instance is translated with respect to its critical circuit depth $p-p^{*}$. Additionally, the energy is shifted according to $F_{p}-E_{\mathrm{OGP}}$. 

\section{Fourier Parameter Schedules}
\label{app:fourier_parameters}
To investigate the evolution of the variational parameters before and after the critical depth $p^{*}$, we analyze the optimized QAOA schedules obtained using the Fourier parameterization introduced by Zhou \textit{et al.}~\cite{Zhou_2020}. Instead of optimizing every QAOA parameter independently, the phase-separation and mixer angles are expanded in a truncated Fourier basis,
\begin{align}
\gamma_i
&=
\sum_{k=1}^{q}
u_k
\sin\!\left[
\left(k-\frac12\right)
\left(i-\frac12\right)
\frac{\pi}{p}
\right],\\
\beta_i
&=
\sum_{k=1}^{q}
v_k
\cos\!\left[
\left(k-\frac12\right)
\left(i-\frac12\right)
\frac{\pi}{p}
\right],
\end{align}
where only the Fourier coefficients $\{u_k,v_k\}$ are optimized. Consequently, the number of optimization variables is reduced from $2p$ to $ 2q,q\ll p$, while maintaining smooth parameter schedules that can be efficiently transferred between different circuit depths and system sizes ~\cite{Zhou_2020}. 

\end{document}